\documentclass[journal]{IEEEtran}
\usepackage{algorithm}
\usepackage{algorithmic}

\usepackage{cite}
\usepackage{multicol,multirow}
\usepackage{makecell,booktabs}
\usepackage{xurl}
\usepackage{hyperref}
\hypersetup{    
    colorlinks=true,
    linkcolor=black,
    anchorcolor=black,
    citecolor=black,
    filecolor=black,
    menucolor=black,
    runcolor=black,
    urlcolor=black,
}

\usepackage{caption}
\usepackage{subcaption} 
\usepackage[utf8]{inputenc} 
\usepackage{graphicx}
\usepackage{amsfonts}
\usepackage{amssymb}
\usepackage{amsmath,mathtools}
\usepackage{array}

\usepackage{color}
\usepackage{svg}
\usepackage{bbm}
\usepackage{bm}
\usepackage{comment}
\usepackage{forest}
\usepackage{etoolbox}
\usepackage{xspace}
\usepackage{kotex}

\usepackage{cleveref}
\usepackage{lipsum}  
\usepackage{balance}
\usepackage{tabularx}
\usepackage{pifont}

\usepackage{siunitx}
\usepackage{verbatim}

\newcolumntype{L}[1]{>{\raggedright\let\newline\\\arraybackslash\hspace{0pt}}m{#1}}
\newcolumntype{C}[1]{>{\centering\let\newline\\\arraybackslash\hspace{0pt}}m{#1}}
\newcolumntype{R}[1]{>{\raggedleft\let\newline\\\arraybackslash\hspace{0pt}}m{#1}}

\usepackage{graphicx} 

\newcommand{\etal}{\textit{et al.}}

\newlength{\origtextfloatsep}
\newlength{\origintextsep}
\newlength{\origfloatsep}

\AtBeginDocument{%
  \setlength{\origtextfloatsep}{\textfloatsep}%
  \setlength{\origintextsep}{\intextsep}%
  \setlength{\origfloatsep}{\floatsep}%
}

\AtBeginEnvironment{algorithm}{%
  \setlength{\textfloatsep}{1pt plus 1pt minus 1pt}
  \setlength{\intextsep}{1pt plus 1pt minus 1pt}
  \setlength{\floatsep}{1pt plus 1pt minus 1pt}
}

\AfterEndEnvironment{algorithm}{%
  \setlength{\textfloatsep}{\origtextfloatsep}%
  \setlength{\intextsep}{\origintextsep}%
  \setlength{\floatsep}{\origfloatsep}%
}

\begin{document}

\author{
    Minsoo~Kim,~\IEEEmembership{Member,~IEEE,}
    Matthew~Brun,
    Andy~Sun,~\IEEEmembership{Senior Member,~IEEE,}
    and Jip~Kim,~\IEEEmembership{Member,~IEEE}
    }

\title{
    {Dispatch-Aware Deep Neural Network for\\Optimal Transmission Switching}
}
\maketitle

\begin{abstract}
    Optimal transmission switching (OTS) improves optimal power flow (OPF) by selectively opening transmission lines, but its mixed-integer formulation increases computational complexity, especially on large grids. To address this, we propose a dispatch-aware deep neural network (DA-DNN) that accelerates DC-OTS without relying on pre-solved labels, eliminating costly OTS label generation that becomes impractical at scale. DA-DNN predicts line states and passes them through an embedded differentiable DC-OPF layer, using the resulting generation cost as the loss function so that physical network constraints are enforced throughout training and inference. To stabilize training, we adopt a customized weight and bias initialization that keeps the embedded DC-OPF feasible from the first epoch. To improve inference robustness, we incorporate a binary regularization term that reduces ambiguity in the relaxed line-status outputs prior to thresholding. Once trained, DA-DNN produces a feasible topology and dispatch pair with highly predictable computation time comparable to a single DC-OPF solve, while conventional MIP solvers can become intractable. Moreover, the embedded OPF layer enables DA-DNN to generalize to untrained system configurations, such as changes in line flow limits, and to support post-contingency corrective operation. As a result, the proposed method captures the economic advantages of OTS while maintaining scalability and generalization ability.
\end{abstract}\vspace{-3mm}

\section*{Selected Nomenclature}\vspace{-2mm}
\addcontentsline{toc}{section}{Nomenclature}
\subsection{Variables}
\begin{IEEEdescription}[\IEEEusemathlabelsep\IEEEsetlabelwidth{$\mathbf{W}^\text{last}, \mathbf{b}^\text{last}$}]
\item[$\mathbf{W}, \mathbf{b}$] Weight and bias of neural network
\item[$\mathbf{p}_g, \mathbf{z}, \bm{\theta}$] Optimization variables
\end{IEEEdescription}\vspace{-4mm}

\subsection{Functions}
\begin{IEEEdescription}[\IEEEusemathlabelsep\IEEEsetlabelwidth{$\mathbf{W}^\text{last}, \mathbf{b}^\text{last}$}]
\item[$\mathcal{L}, L, B$] Loss/Lagrange/binarization function
\item[$C, R$] Generation cost/binary regularization function
\end{IEEEdescription}\vspace{-4mm}

\subsection{Parameters}
\begin{IEEEdescription}[\IEEEusemathlabelsep\IEEEsetlabelwidth{$\mathbf{W}^\text{last}, \mathbf{b}^\text{last}$}]
\item[$\mathbf{M}, \mathbf{C}$] Generator-bus/branch-bus incidence matrix
\item[$\overline{\mathbf{p}}_g, \underline{\mathbf{p}}_g, \overline{\bm{\theta}}, \underline{\bm{\theta}}$] Generation/voltage angle limits
\item[$\overline{\mathbf{p}}_l, \underline{\mathbf{p}}_l, \mathbf{b}_l$] Line flow limits/line susceptance
\end{IEEEdescription}

\section{Introduction}

Optimal transmission switching (OTS) strategically opens or closes transmission lines so that the network topology becomes an active control variable \cite{fisher2008optimal}. By doing so, system operators can reroute power flows and reduce generation costs. This effect is known as Braess’s paradox in power systems, where removing a line can redirect flows and reduce the total operational cost \cite{schafer2022understanding}. Beyond this counterintuitive phenomenon, OTS provides a practical mechanism for alleviating congestion, accessing cheaper generation, and improving system flexibility without building new infrastructure. This is because treating the line status as a decision variable enlarges the feasible set and enables more cost-effective dispatch solutions from an optimization perspective.

Studies show that transmission switching can deliver significant economic and reliability improvements. For example, simulations from the ARPA-E Topology Control Algorithm project indicate that transmission switching in the PJM system can reduce congestion costs by more than 50\% and save over 100 million USD annually in the real-time market \cite{ARPA_E2014}. In addition, ERCOT has used transmission switching to replace post-contingency load shedding and to support more effective corrective and preventive operating plans \cite{ruiz2020topology}. In MISO, one reconfiguration applied during a 345-kV outage reduced regional wind curtailments by 86\% and reduced congestion costs by 3.5 million USD \cite{RuizMyhre2024MISO}. In SPP, a real-time operational case study found that even a single switching action could reduce the power flow on a congested transmission line by more than 25\% \cite{ruiz2020topology}. 

Recognizing the operational benefits of transmission switching, regulatory and policy have begun to encourage its evaluation alongside other grid-enhancing technologies. For example, FERC Order No. 1920 in the United States and ENTSO-E’s Research and Innovation Roadmap 2017–2026 in Europe explicitly identify transmission switching as an option to improve grid flexibility and operational efficiency \cite{fercorder1920, entsoe2017roadmap}. In parallel with this regulatory momentum, system operators such as PJM and ISO-NE already allow limited switching in response to contingencies to alleviate overloads and maintain reliability \cite{PJMManual37, ISONEOP19}.

Although there is a huge potential in OTS, its adoption in practice has been hindered by high computational complexity \cite{numan2023role}. Fully exploiting the value of transmission switching requires obtaining globally optimal transmission switching decisions by introducing binary line status of optimization variables. By doing so, the optimal power flow (OPF) problem is transformed into a mixed-integer program (MIP) that is NP hard and often intractable for large networks \cite{crozier2022feasible}. Commercial solvers may take minutes to days or fail to converge entirely, which prevents OTS from being used in both day-ahead and real-time operations. Thus, faster OTS solution methods are essential, and a variety of approaches have been explored, as summarized in Table~\ref{table:related_works_ots}.

One major direction is the use of heuristic algorithms that construct high quality switching decisions without exhaustively solving the full MIP. Early works on OTS focused on heuristic approaches that reduce computation by narrowing the search over switching actions. These methods typically rely on the structural features of the power system or simplified representations of problems. For example, sensitivity-based methods rank candidate lines using power transfer distribution factors or shadow prices \cite{ruiz2012tractable}, while other studies rely on OPF-derived congestion indicators \cite{fuller2012fast}. Greedy heuristic algorithms iteratively open lines to reduce cost, as in \cite{hedman2009optimal}. Topological metrics have been tested as heuristics for screening candidate lines \cite{barrows2013using}. Beyond heuristic screening, approximate OTS formulations modify the MIP structure to implicitly penalize switching, enabling near-optimal dispatch while producing much simpler topologies and reducing computation time \cite{jabarnejad2018approximate}. Although these methods can yield feasible decisions, they still require solving numerous OPF or OTS subproblems. These limitations motivate learning-based approaches that aim to better capture the complexity of OTS.

\begin{table}[t]
    \centering
    \captionsetup{justification=centering, labelsep=period, font=footnotesize, textfont=sc}
    \caption{Comparison of OTS methodologies.}
    \label{table:related_works_ots}
    \begin{tabular}{l|ccc}
        \toprule
        \makecell{Work}
        & \makecell{Method} 
        & \makecell{Feasibility\\of dispatch}
        & \makecell{Evaluation under\\untrained constraints}\\
        \midrule\midrule
        Ruiz \etal \cite{ruiz2012tractable}
          & Heuristic
          & \checkmark 
          & $\times$ \\
        Fuller \etal \cite{fuller2012fast}
          & Heuristic
          & \checkmark 
          & $\times$ \\
        Hedman \etal \cite{hedman2009optimal}
          & Heuristic
          & \checkmark 
          & $\times$ \\
        Barrows \etal \cite{barrows2013using}
          & Heuristic
          & \checkmark 
          & $\times$ \\
        Jabarnejad \etal \cite{jabarnejad2018approximate}
          & Heuristic
          & \checkmark 
          & $\times$ \\
          \midrule
        Han \etal \cite{han2022learning}
          & RL
          & $\times$
          & $\times$ \\
        Lehna \etal \cite{lehna2023managing}
          & RL
          & $\times$ 
          & $\times$ \\
        Tang \etal \cite{tang2022optimal}
        & RL
          & $\times$
          & $\times$ \\
        Pineda \etal \cite{pineda2024learning}
          & SL
          & $\times$
          & $\times$ \\
        Yang \etal \cite{yang2019line}
        & SL
          & $\times$ 
          & $\times$ \\
        Bugaje \etal \cite{bugaje2023real}
          & SL
          & \checkmark
          & $\times$ \\
        \midrule
        \textbf{DA-DNN (proposed)}
          & USL
          & \textbf{\checkmark}
          & \checkmark \\
        \bottomrule
    \end{tabular}
    \captionsetup{justification=raggedright, singlelinecheck=false, textfont=normalfont}
    \caption*{RL/SL/USL: reinforcement/supervised/unsupervised learning.}\vspace{-7mm}
\end{table}

Although learning-based approaches have been explored as an alternative to heuristic OTS methods, there are several challenges. \textbf{First}, many learning approaches rely on supervised learning, which requires pre-solved OTS as labels. Obtaining these labels requires repeatedly solving the OTS problem itself, which is computationally expensive and often infeasible for large networks, thereby eliminating the intended advantage of learning-based acceleration. \textbf{Second}, end-to-end learning methods that directly predict switching decisions or dispatch values do not enforce power balance, line limits, or generator limits, and can therefore produce infeasible operating points. This is a critical challenge in power system operations, where feasibility violations can lead to insecure or unsafe dispatch outcomes. \textbf{Third}, existing methods lack adaptability to changes in network constraints. In practical system operations, line flow limits often change due to seasonal conditions or technologies such as dynamic line rating. As a result, models must remain reliable under varying constraint settings to ensure safe and efficient operation.

There have been several efforts in the literature to address these challenges. Several works have attempted to address the first challenge by adopting reinforcement learning (RL) to learn switching strategies directly from interaction with a simulated grid environment. For example, DDQN agents were employed to sequentially remove branches without requiring pre-solved optimal solutions \cite{han2022learning}, and RL agents have been applied to real-time topology control to manage overloads and instability \cite{lehna2023managing}. RL has also been explored for OTS applications such as short-circuit current mitigation, where DQN agents choose switching actions-based on reward feedback \cite{tang2022optimal}. Although RL removes label dependence, it does not guarantee feasibility. This is because it optimizes for rewards instead of enforcing physical constraints of the power system, and can therefore produce infeasible or unsafe switchings.

To address the second challenge, some studies incorporate feasibility mechanisms during inference. One common approach is to combine learning with optimization by warm starting a solver using predicted switching actions, allowing feasibility to be enforced through a conventional OPF or MIP solver \cite{pineda2024learning}. Other methods apply feasibility checks followed by OPF-based correction steps if the predicted topology leads to constraint violations \cite{bugaje2023real}. However, the performance after post-processing is not guaranteed, as reported in \cite{kim2022projection}. Furthermore, these methods typically rely on labeled data during training, which involves significant computational overhead to prepare. Machine learning approaches for OTS, such as line-ranking and algorithm-selection methods \cite{yang2019line}, also require labeled optimal switching decisions. Thus, these works do not resolve the first challenge about label dependency.

Although existing methods address either label dependency or feasibility, none explicitly tackle the third challenge. Most prior work evaluates learning models under fixed network constraints, without considering variability in system parameters. As a result, their performance can significantly degrade or remain unchanged when constraint settings change. This lack of adaptability limits their applicability in real-world system operations, where constraint variability is common \cite{wallnerstrom2014impact}. Although learning-based OPF methods have investigated generalization under varying network information \cite{liu2022topology}, no studies have validated this capability in the context of OTS.

In this regard, we propose a dispatch aware deep neural network (DA-DNN) to overcome the aforementioned three challenges of learning-based OTS. To address the first challenge, DA-DNN is trained in an unsupervised manner using the generation cost from a differentiable DC-OPF layer as the loss function, eliminating the need for labeled OTS solutions. For the second challenge, the embedded DC-OPF layer enforces dispatch feasibility given the predicted topology during both training and inference. Finally, the third challenge is resolved by incorporating the optimization layer directly into the model. Note that the parameters of the model are updated through implicit differentiation, which is adopted in the learning-based OPF \cite{kim2025mpa}. This enables DA-DNN to remain adaptive under changes in network constraints.

This paper builds on our preliminary work in \cite{kim2025dispatch} and makes the following key contributions:

\begin{enumerate}
    \item We propose a novel DA-DNN to accelerate solving DC-OTS which consists of a line switching network and a DC-OPF layer. During training, the line switching network outputs predicted line statuses. Next, the embedded DC-OPF layer solves DC-OPF with these line statuses to obtain the generator dispatch. The resulting generation cost is used as an unsupervised loss, which eliminates the need for precomputed OTS as labels. Since every forward pass solves a DC-OPF, all physical constraints are enforced throughout training and inference. After training, we binarize the relaxed line status and run DC-OPF to obtain a pair of line status and dispatch that is  feasible. As a result, the total computation time for our method in the inference phase is always equal to that of a single DC-OPF. To the best of our knowledge, this is the first work that solves DC-OTS in an unsupervised learning method with an embedded DC-OPF layer.

    \item We propose two complementary mechanisms to stabilize training and improve inference performance. First, we introduce a weight and bias initialization that makes the embedded DC-OPF feasible at the first epoch, enabling stable training where standard He initialization \cite{he2015delving} suffers from infeasible solutions. Second, we incorporate a binary regularization term that reduces ambiguity in the predicted relaxed line statuses, resulting in more reliable binarized switchings and an improved generation cost during inference. Together, these mechanisms ensure that training proceeds from a feasible operating region while guiding the relaxed line statuses toward stable binary decisions, thereby enhancing both learning stability and post-thresholding performance. We further benchmark against a solver-based continuous relaxation with binarization, showing its feasibility degradation and highlighting the advantage of learning the relaxed topology through the differentiable DC-OPF layer.

    \item We evaluate the proposed framework on the IEEE 73, 118, and 300 bus systems. The results show that each inference requires the same computation time as a single DC-OPF solve. In the 300-bus case, where a commercial solver fails to complete DC-OTS within days, the proposed method produces feasible line statuses that reduce generation cost within milliseconds. The method further demonstrates robustness to untrained line flow limits and applicability to post-contingency corrective operation.
    
\end{enumerate}

\section{DC Optimal Transmission Switching Model}

Suppose that the given power system consists of $N_b$ buses, $N_g$ generators, and $N_l$ lines. Let $\Xi_\text{OTS}:=\{\mathbf{p}_g, \bm{\theta}, \mathbf{z}\}$ be the set of optimization variables, where $\mathbf{p}_g\in\mathbb{R}^{N_g}$, $\bm{\theta}\in\mathbb{R}^{N_b}$, and $\mathbf{z}\in\{0,1\}^{N_l}$ denote active power generation, voltage phase angle, and line on/off status. Then, the DC-OTS problem is formulated as follows \cite{ahmadi2023decomposition}:
\begingroup
\allowdisplaybreaks
\begin{subequations}\label{Eq:DC-OTS}
\begin{align}
&
\min_{\Xi_\text{OTS}}
C(\mathbf{p}_g)
\label{Eq:OTS_Obj}\\
&
\mathbf{M}\mathbf{p}_g - \mathbf{p}_d = \mathbf{C}^\intercal \big(\text{diag}(\mathbf{z} \odot \mathbf{b}_l)\mathbf{C}\bm{\theta}\big),\label{Eq:OTS_PF}\\
&
\mathbf{z}\odot\underline{\mathbf{p}}_l \leq \text{diag}(\mathbf{z} \odot \mathbf{b}_l)\mathbf{C}\bm{\theta} \leq \mathbf{z}\odot\overline{\mathbf{p}}_l,\label{Eq:OTS_line_limits}\\
&
\underline{\mathbf{p}}_g \leq \mathbf{p}_g \leq \overline{\mathbf{p}}_g,\label{Eq:OTS_gen_limits}\\
&
\underline{\bm{\theta}}\leq \bm{\theta} \leq \overline{\bm{\theta}},\label{Eq:OTS_angle_limits}\\
&
\theta_\text{ref} = 0, \label{Eq:OTS_slack}
\end{align}
\end{subequations}
\endgroup
\noindent 
where $C(\cdot):\mathbb{R}^{N_g}\rightarrow\mathbb{R}$ in (\ref{Eq:OTS_Obj}) is the convex generation cost function of each generator. (\ref{Eq:OTS_PF}) enforces DC power balance equation, where $\mathbf{M}\in \{0,1\}^{N_b \times N_g}$ is generator-bus incidence matrix that maps $\mathbf{p}_g$ to the vectors of $N_b$ dimensions, $\mathbf{p}_d\in\mathbb{R}^{N_b}$ denotes nodal demand of each bus, and $\text{diag}(\mathbf{z}\odot\mathbf{b}_l)\mathbf{C}\bm{\theta}$ represents the line flows, which are restricted by $\underline{\mathbf{p}}_l$ and $\overline{\mathbf{p}}_l$ in (\ref{Eq:OTS_line_limits}). Here, $\mathbf{C}\in\{-1,0,1\}^{N_l\times N_b}$ is the branch-bus incidence matrix, $\mathbf{b}_l\in\mathbb{R}^{N_l}$ is the vector of line susceptances, and $\odot$ refers to element-wise multiplication. Note that $\mathbf{z}$ nullifies any line flows that are switched out. In addition, the output of the generator and the angle of the nodal voltage are limited by (\ref{Eq:OTS_gen_limits}) and (\ref{Eq:OTS_angle_limits}), respectively. Finally, (\ref{Eq:OTS_slack}) defines the slack bus.

\begin{figure*}[t]
	\centering
\includegraphics[width=1.0\textwidth]{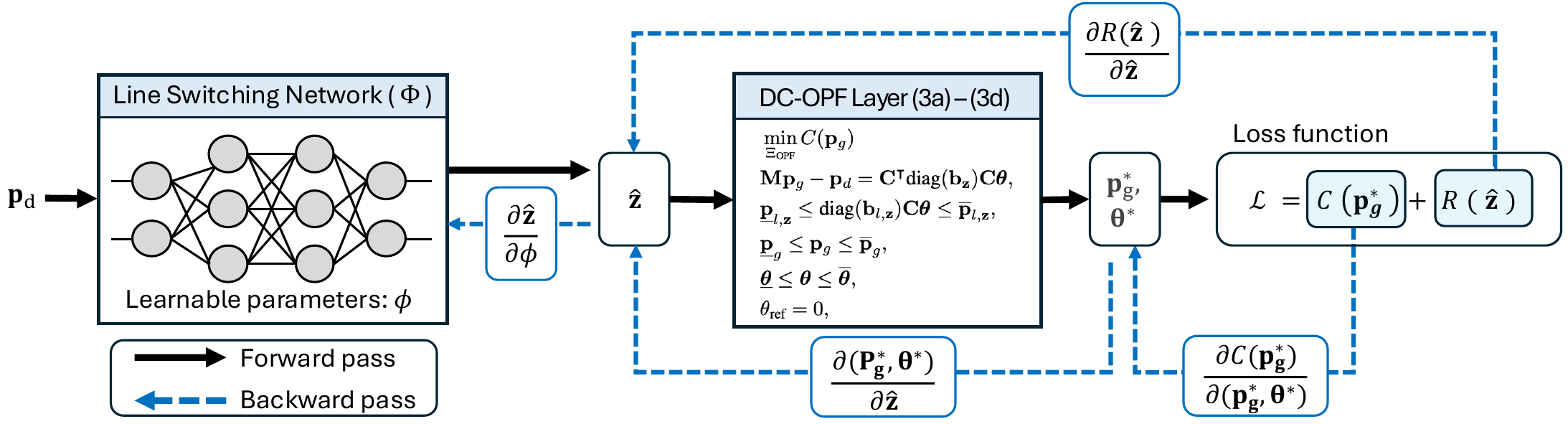}\vspace{-2mm}
	\caption{Training process of the proposed DA-DNN for optimal transmission switching.}
	\label{fig:DA-DNN_framework}\vspace{-5mm}
\end{figure*}

\section{Methodologies}

Unlike DC-OPF, DC-OTS is NP-hard due to the binary vector $\mathbf{z}$. Thus, commercial MIP solvers may require several hours or even days to reach optimality. Consequently, practical adoption hinges on accelerated solution techniques. Thus, in this section, we introduce the proposed DA-DNN, an efficient DNN-based method that obtains cost-effective switching decisions within operational time limits. Importantly, the OPF layer enforces the feasibility of dispatch, which prevents infeasible dispatch outcomes during training and inference.

The overall framework of the proposed DA-DNN is illustrated in Fig.~\ref{fig:DA-DNN_framework}. As shown in the figure, DA-DNN consists of a line switching network and a DC-OPF layer. The line switching network takes bus demand and predicts the line status of the network. Then, DC-OPF layer takes the predicted line status, solves DC-OPF based on that values, and obtains dispatch. After that, the model learns to reduce the generation cost that is determined from the obtained dispatch. Note that pre-solved labels of DC-OTS are not required when training the proposed DA-DNN.

\subsection{Forward Pass of DA-DNN}

\subsubsection{Solving DC-OPF with NN predicted line status}
Let $\Phi:\mathbb{R}^{N_b}\rightarrow\mathbb{R}^{N_l}$ be the line switching network in Fig.~\ref{fig:DA-DNN_framework} that maps $\mathbf{p}_d$ to the predicted line switching state $\mathbf{\hat{z}}\in [0, 1]^{N_l}$, thus we have $\mathbf{\hat{z}} = \Phi(\mathbf{p}_d)$. Note that all elements of $\mathbf{\hat{z}}$ are bounded by $[0,1]$, since $\mathbf{\hat{z}}$ is used as a continuous relaxation of the binary line switch state when solving (\ref{Eq:DC-OTS_relax}). For this bound, we use a sigmoid function $\sigma(\cdot)$ as follows:
\begin{equation}
    \mathbf{\hat{z}} = \sigma\Big(\mathbf{W}^\text{last}\mathbf{\tilde{z}} + \mathbf{b}^\text{last}\Big)
\end{equation}
where $\mathbf{W}^\text{last}\in\mathbb{R}^{N_l \times N_h}$, $\mathbf{b}^\text{last}\in\mathbb{R}^{N_l}$, and $\mathbf{\tilde{z}}$ are the weight matrix, bias vector, and input of the last layer of $\Phi$. Note that $N_h$ is the dimension of hidden vector.

Now, we solve the DC-OPF using $\mathbf{\hat{z}}$. Let $\Xi_\text{OPF} := \{\mathbf{p}_g, \bm{\theta} \}$ be the set of optimization variables of DC-OPF. Then, the optimization problem is formulated as follows:
\begingroup
\allowdisplaybreaks
\begin{subequations}\label{Eq:DC-OTS_relax}
\begin{align}
&
\min_{\Xi_\text{OPF}}
C(\mathbf{p}_g)
\label{Eq:OTS_z_Obj}\\
&
\mathbf{M}\mathbf{p}_g - \mathbf{p}_d = \mathbf{C}^\intercal \text{diag}(\mathbf{b}_{l,\hat{\mathbf{z}}})\mathbf{C}\bm{\theta},\label{Eq:OTS_z_PF}\\
&
\underline{\mathbf{p}}_{l,\hat{\mathbf{z}}} \leq\text{diag}(\mathbf{b}_{l,\hat{\mathbf{z}}})\mathbf{C}\bm{\theta} \leq \overline{\mathbf{p}}_{l,\hat{\mathbf{z}}},\label{Eq:OTS_z_line_limits}\\
&
\text{(\ref{Eq:OTS_gen_limits})--(\ref{Eq:OTS_slack})},\label{Eq:OPF_same_constraints}
\end{align}
\end{subequations}
\endgroup
\noindent
where $\mathbf{b}_{l,\hat{\mathbf{z}}} = \hat{\mathbf{z}}\odot\mathbf{b}_l$, $\underline{\mathbf{p}}_{l,\hat{\mathbf{z}}} = \hat{\mathbf{z}}\odot\underline{\mathbf{p}}_{l}$, and $\overline{\mathbf{p}}_{l,\hat{\mathbf{z}}} = \hat{\mathbf{z}}\odot\overline{\mathbf{p}}_{l}$. This optimization problem is DC-OPF with a fixed continuously relaxed line switching state $\mathbf{\hat{z}}$. 

\subsubsection{Weight and bias initialization}\label{subsubsec:weight_bias_init}
In the initial training process, (\ref{Eq:DC-OTS_relax}) is likely infeasible, because switching decisions from an untrained NN are arbitrary and distort or disconnect the network. Thus, training of the proposed DA-DNN becomes unstable and cannot progress.

Let \(\mathbf W_{\mathrm{init}}^\text{last}\) and \(\mathbf b_{\mathrm{init}}^\text{last}\) denote the initial weight matrix and the bias vector of the last layer of \(\Phi\), respectively.  
The resulting initial relaxed line state $\hat{\mathbf{z}}_\text{init}$ is
\begin{equation}
    \hat{\mathbf z}_{\mathrm{init}}
  =\sigma\!\Bigl(\mathbf W_{\mathrm{init}}^\text{last}\tilde{\mathbf z}_{\mathrm{init}}+\mathbf b_{\mathrm{init}}^\text{last}\Bigr),
\end{equation}
where $\tilde{\mathbf z}_{\mathrm{init}}$ is the input to the last initialized layer. We set \(\mathbf W_{\mathrm{init}}^\text{last}=\mathbf 0\) and \(\mathbf b_{\mathrm{init}}^\text{last}=9\), so that every component is evaluated at $\hat{\mathbf{z}}_\text{init} = \sigma(9) = 0.9999$. Employing this $\hat{\mathbf z}_{\mathrm{init}}$ for~\eqref{Eq:DC-OTS_relax} solves the optimization problem with the standard DC-OPF. 

\subsubsection{Loss function to train the model}
Let $\Xi_\text{OPF}^* := \{\hat{\mathbf{p}}_g^*, \hat{\bm{\theta}}^*\}$ denote the optimal solution of the relaxed DC-OPF~\eqref{Eq:DC-OTS_relax} associated with the predicted line status $\hat{\mathbf{z}}$, defined as 
\begin{equation}
    \Xi_\text{OPF}^*
    := \{\mathbf{\hat{p}}_g^*, \bm{\hat{\theta}}^*\}
    = \arg\min_{\Xi_\text{OPF}\in\mathcal{F}(\hat{\mathbf{z}})} C(\mathbf{p}_g),
    \label{Eq:argmin_function}
\end{equation}
where $\mathcal{F}(\cdot)$ denotes the feasible region.
Since the DC-OPF layer explicitly enforces all physical constraints, the resulting dispatch is always feasible.

However, transmission switching decisions must be binary. Accordingly, in the inference phase, the relaxed line status $\hat{\mathbf{z}}$ is converted into a binary topology using a fixed threshold (e.g., $0.5$), and a single DC-OPF is solved under the resulting topology to obtain the final feasible dispatch. To ensure that this thresholding operation is robust, we train DA-DNN using the following unsupervised loss function:
\begin{equation}
    \mathcal{L}
    = C(\hat{\mathbf{p}}_g^*)
    + R(\hat{\mathbf{z}}).
    \label{Eq:loss_function}
\end{equation}
The first term of (\ref{Eq:loss_function}) directly minimizes the generation cost obtained from the embedded DC-OPF layer, thereby guiding the predicted line status toward topologies that reduce the overall generation cost. The second term is a binary regularization that penalizes ambiguous line statuses, which is defined as $R(\hat{\mathbf{z}}) = ||\alpha\hat{\mathbf{z}}\odot(1-\hat{\mathbf{z}})||_1$ where $\alpha\geq0$ \cite{zhang2007binary}. Although continuous relaxation of $\mathbf{z}$ is required to enable differentiation through the DC-OPF layer during training, minimizing $R(\hat{\mathbf{z}})$ encourages $\hat{\mathbf{z}}$ to approach $0$ or $1$. This reduces the discrepancy between the relaxed solutions used during training and the binary line statuses applied during inference.

\subsection{Backpropagation Through DC-OPF Layer}

The backpropagation to train $\Phi$ with learnable parameters $\bm{\phi}$ (e.g., $\mathbf{W}$ and $\mathbf{b}$) is expressed as follows:
\begin{equation}
\begin{aligned}
    \nabla_{\bm{\phi}} \mathcal{L} &= \frac{\partial C(\hat{\mathbf{p}}_g^*)}{\partial \bm{\phi}} + \frac{\partial R(\hat{\mathbf{z}})}{\partial \bm{\phi}}\\
    &= \Bigg(\frac{\partial C(\hat{\mathbf{p}}_g^*)}{\partial (\mathbf{\hat{p}}_g^*, \mathbf{\hat{\theta}}^*)}\frac{\partial (\mathbf{\hat{p}}_g^*, \mathbf{\hat{\theta}}^*)}{\partial \mathbf{\hat{z}}} + \frac{\partial R(\hat{\mathbf{z}})}{\partial \hat{\mathbf{z}}}\Bigg)\frac{\partial \mathbf{\hat{z}}}{\partial \bm{\phi}}.\label{Eq:back_propagation}
\end{aligned}
\end{equation}
Here, $\frac{\partial C(\hat{\mathbf{p}}_g^*)}{\partial (\mathbf{\hat{p}}_g^*, \mathbf{\hat{\theta}}^*)}$ and $\frac{\partial R(\hat{\mathbf{z}})}{\partial \hat{\mathbf{z}}}$ are determined analytically.  $\frac{\partial \mathbf{\hat{z}}}{\partial \bm{\phi}}$ is determined from automatic differentiation, which is widely adopted to train deep learning models. Thus, we need to obtain $\frac{\partial (\mathbf{\hat{p}}_g^*, \mathbf{\hat{\theta}}^*)}{\partial \mathbf{\hat{z}}}$, which requires derivatives of the argmin function (\ref{Eq:argmin_function}).

For this, we simplify the DC-OPF problem (\ref{Eq:DC-OTS_relax}) as follows:
\begingroup
\allowdisplaybreaks
\begin{subequations}\label{Eq:simple_OPF}
\begin{align}
&
\min_{\mathbf{x}}
f(\mathbf{x})
\label{Eq:simple_obj}\\
&
g(\mathbf{x},  \mathbf{\hat{z}}) = 0,\label{Eq:simple_PF}\\
&
h(\mathbf{x},  \mathbf{\hat{z}}) \leq 0,\label{Eq:simple_limits}
\end{align}
\end{subequations}
\endgroup
\noindent
where $\mathbf{x}$ denotes the optimization variables of (\ref{Eq:DC-OTS_relax}). Then, the Lagrange function $L$ is formulated as:
\begin{equation}
    L(\mathbf{x}, \bm{\lambda}, \bm{\mu}; \mathbf{\hat{z}}) = f(\mathbf{x}) + \bm{\lambda}^\intercal g(\mathbf{x},  \mathbf{\hat{z}}) + \bm{\mu}^\intercal h(\mathbf{x},  \mathbf{\hat{z}})
\end{equation}
where $\bm{\lambda}$ and $\bm{\mu}\geq 0$ are the Lagrange multipliers of equality and inequality constraints. Since (\ref{Eq:simple_OPF}) is a convex optimization problem, the KKT conditions are necessary and, under Slater’s condition, sufficient for optimality. Thus, we have
\begin{equation}
    \mathcal{I} := \begin{bmatrix}
        \frac{\partial L(\mathbf{x}, \bm{\lambda}, \bm{\mu}; \mathbf{\hat{z}})}{\partial \mathbf{x}}\\
        g(\mathbf{x}, \mathbf{\hat{z}})\\
        \bm{\mu}\odot h(\mathbf{x}, \mathbf{\hat{z}})
        \end{bmatrix}  =  \mathbf{0}
\end{equation}
where $\mathcal{I}$ is the implicit function of $\mathbf{x}$ and $\mathbf{\hat{z}}$ for the KKT optimality conditions. Since the derivative of $\mathcal{I}$ with respect to $\mathbf{\hat{z}}$ exists due to the implicit function theorem \cite{krantz2002implicit}, we have
\begin{equation}
    \frac{\partial \mathcal{I}}{\partial \mathbf{\hat{z}}} + \frac{\partial \mathcal{I}}{\partial \mathbf{x}}\frac{\partial \mathbf{x}}{\partial \mathbf{\hat{z}}} = \mathbf{0}.
\end{equation}
Since $\mathbf{x}$ represents $\mathbf{\hat{p}}_g^*$ and $ \mathbf{\hat{\theta}}^*$, we have
\begin{equation}
    \frac{\partial (\mathbf{\hat{p}}_g^*, \mathbf{\hat{\theta}}^*)}{\partial \mathbf{\hat{z}}} = -\bigg(\frac{\partial \mathcal{I}}{\partial (\mathbf{\hat{p}}_g^*, \mathbf{\hat{\theta}}^*)}\bigg)^{-1}\frac{\partial \mathcal{I}}{\partial \mathbf{\hat{z}}}.
\end{equation}
It should be noted that $\frac{\partial \mathcal{I}}{\partial \mathbf{\hat{z}}}$ and $\frac{\partial \mathcal{I}}{\partial \mathbf{x}}$ are both explicitly determined. Hence, we now have the backpropagation process to update $\bm{\phi}$ of $\Phi$ as follows:
\begin{equation}
    \nabla_{\bm{\phi}} \mathcal{L} = \Bigg(-\frac{\partial \mathcal{L}}{\partial (\mathbf{\hat{p}}_g^*, \mathbf{\hat{\theta}}^*)}\bigg(\frac{\partial \mathcal{I}}{\partial (\mathbf{\hat{p}}_g^*, \mathbf{\hat{\theta}}^*)}\bigg)^{-1}\frac{\partial \mathcal{I}}{\partial \mathbf{\hat{z}}} + \frac{\partial R(\hat{\mathbf{z}})}{\partial \hat{\mathbf{z}}}\Bigg) \frac{\partial \mathbf{\hat{z}}}{\partial \bm{\phi}}.\label{Eq:back_propagation_implict}
\end{equation}
Thus, the learnable parameter $\bm{\phi}$ is updated as follows:
\begin{equation}
    \bm{\phi} \leftarrow \bm{\phi} - \eta_\text{lr}\Bigg(-\frac{\partial \mathcal{L}}{\partial (\mathbf{\hat{p}}_g^*, \mathbf{\hat{\theta}}^*)}\bigg(\frac{\partial \mathcal{I}}{\partial (\mathbf{\hat{p}}_g^*, \mathbf{\hat{\theta}}^*)}\bigg)^{-1}\frac{\partial \mathcal{I}}{\partial \mathbf{\hat{z}}} + \frac{\partial R(\hat{\mathbf{z}})}{\partial \hat{\mathbf{z}}}\Bigg) \frac{\partial \mathbf{\hat{z}}}{\partial \bm{\phi}}
\end{equation}
where $\eta_\text{lr}$ denotes the learning rate of DA-DNN. We provide the training process of the proposed method in Algorithm~\ref{alg:DADNN}.

\begin{algorithm}[t]
\caption{Training process of DA-DNN}\label{alg:DADNN}
\begin{algorithmic}[1]
\STATE \textbf{Inputs:}\\
Training dataset $\mathcal{D}=\{\mathbf{p}_d^{(n)}\}_{n=1}^{N}$, Learning rate $\eta_{lr}$,\\
Maximum training epochs $T$, DC-OPF solver $\mathcal{S}_{\text{OPF}}(\cdot)$,\\
NN parameters $\phi = (\mathbf{W}, \mathbf{b})$.\\
\STATE \textbf{Initialize:} \\
Initialize $\mathbf{W}$ and $\mathbf{b}$, Set $\mathbf{W}_{\text{last}}^\text{init} = \mathbf{0}$, $\mathbf{b}_{\text{last}}^\text{init} = 9$
\FOR{$t \in \{1, ..., T\}$}
    \FORALL{$\mathbf{p}_d \in \mathcal{D}$}
        \STATE $\hat{\mathbf{z}} = \Phi(\mathbf{p}_d)$ \hfill
        \STATE $(\hat{\mathbf{p}}_g^*, 
        \hat{\boldsymbol{\theta}}^*) \gets \mathcal{S}_{\text{OPF}}(\hat{\mathbf{z}}, \mathbf{p}_d)$
        \STATE $\mathcal{L} = C(\hat{\mathbf{p}}_g^*) + R(\hat{\mathbf{z}})$ \hfill
        \STATE $\boldsymbol{\phi} \leftarrow \boldsymbol{\phi} - \eta_\text{lr}\nabla_{\boldsymbol{\phi}}\mathcal{L}$
    \ENDFOR
\ENDFOR
\STATE \textbf{Output:} $\phi^*$
\end{algorithmic}
\end{algorithm}

\subsection{Inference Process of DA-DNN}

In the inference phase, DA-DNN uses binarized line-status
decisions rather than the continuous relaxed outputs used
during training. Given a load vector, the model first produces
the relaxed switching vector $\hat{\mathbf{z}} = [\hat{z}_i]_{i=1}^{N_l} \in [0,1]^{N_l}$,
which is converted into a binary topology by applying a fixed
threshold:
\begin{equation}
    \bar{z}_i = B(\hat{z}_i) =
    \begin{cases}
    1, & \hat{z}_i \geq 0.5,\\[2pt]
    0, & \hat{z}_i < 0.5, 
    \end{cases}\label{eq:binarization}
\end{equation}
where $\mathbf{\bar{z}} = [\bar{z}_i]_{i=1}^{N_l}\in\{0,1\}^{N_l}$. A single DC-OPF is then solved under this fixed topology to obtain the final dispatch. The inference procedure
is summarized in Algorithm~\ref{alg:DADNN_inference}.

\begin{algorithm}[t]
\caption{Inference process of DA-DNN}
\label{alg:DADNN_inference}
\begin{algorithmic}[1]
\STATE \textbf{Inputs:}\\
Inference dataset $\mathcal{D}_{\text{inf}} = \{\mathbf{p}_d^{(n)}\}_{n=1}^{N_{\text{inf}}}$, Binarization function $B(\cdot)$,
Solver $\mathcal{S}_{\text{OPF}}(\cdot)$, Trained parameters $\boldsymbol{\phi}^* = (\mathbf{W}^*, \mathbf{b}^*)$. \\
\STATE \textbf{Initialize:}\\
Set $\boldsymbol{\phi} \leftarrow \boldsymbol{\phi}^*$
\FORALL{$\mathbf{p}_d \in \mathcal{D}_{\text{inf}}$}
    \STATE $\bar{\mathbf{z}} = B(\Phi(\mathbf{p}_d))$
    \STATE $(\hat{\mathbf{p}}_g^*, \hat{\boldsymbol{\theta}}^*) 
      \gets \mathcal{S}_{\text{OPF}}(\bar{\mathbf{z}}, \mathbf{p}_d)$
\ENDFOR

\STATE \textbf{Output:} A set of $(\mathbf{p}_d, \bar{\mathbf{z}}, \hat{\mathbf{p}}_g^*, 
      \hat{\boldsymbol{\theta}}^*)$.
\end{algorithmic}
\end{algorithm}

\section{Case Study}

\subsection{Test Systems and Data Generation}
\label{subsec:test_system}

We evaluate the proposed DA-DNN on test systems from the PGLib-OPF library~\cite{babaeinejadsarookolaee2019power}. We consider the IEEE 73, 118, and 300 bus systems for evaluation.

For each test system, we generate a dataset of load snapshots by uniformly scaling the base-case demands. For example, let $\mathbf{p}_d^{\text{base}} \in \mathbb{R}^{N_b}$ denote the nominal bus demands from PGLib. For each data $n$, we draw a scalar factor $\mathbf{\kappa}^{(n)} = [\kappa_i^{(n)}]_{i=1}^{N_b}$ where $\kappa_i^{(n)}~\sim~\mathcal{U}(0.90, 1.10)$ and construct $\mathbf{p}_d^{(n)} = \kappa^{(n)} \odot \mathbf{p}_d^{\text{base}}$. In this way, each scenario corresponds to a different bus-wise loading level between 90\% and 110\% of the base case. We retain only scenarios that are at least feasible for DC-OPF and continue this process until we obtain 3,000 feasible load data. For each data, the feasible load snapshots are randomly partitioned into three disjoint sets: 50\% for training, 16.7\% for validation, and 33.3\% for testing. Generator cost coefficients, capacity limits, and line limits are determined from the corresponding PGLib test system. Voltage angle limits are set to 0.6, as done in  \cite{fisher2008optimal, crozier2022feasible}.

\subsection{Baselines and Implementations Details}

\subsubsection{Baselines} We compare the proposed DA-DNN against  conventional optimization and learning-based baselines:

\begin{table*}[t]
	\centering
    \captionsetup{justification=centering, labelsep=period, font=footnotesize, textfont=sc}    
	\caption{Overall Performance Comparisons. \textbf{Bold} indicates the lowest generation cost among learning-based methods with no constraint violations. \textcolor{red}{red} indicates constraint violations and high computation time. We use 200 samples for 300 bus system.}

	\label{table:118_73_bus}
	\begin{tabular}{l|c|cccc|cccc|cccc}
		\toprule
        \makecell[c]{\multirow{6}{*}{{Problem}}}&\multirow{6}{*}{\makecell[c]{Method}}
&\multicolumn{4}{c|}{\makecell{IEEE 73 bus system}}&\multicolumn{4}{c|}{\makecell{IEEE 118 bus system}}&\multicolumn{4}{c}{\makecell{IEEE 300 bus system}}
        \\\cmidrule{3-14}
        & & \makecell{Avg.\\gen.\\cost\\(\$1k)} & \makecell{Avg.\\ineq.\\viol.\\(\%)} &\makecell{Avg\\ eq.\\viol.\\(\%)}  &\makecell{Avg.\\comp.\\ time\\(sec)}&\makecell{Avg.\\gen.\\cost\\(\$1k)} & \makecell{Avg.\\ineq.\\viol.\\(\%)} &\makecell{Avg.\\eq.\\viol.\\(\%)}  &\makecell{Avg.\\comp.\\time\\(sec)}&\makecell{Avg.\\gen.\\cost\\(\$1k)} & \makecell{Avg.\\ineq.\\viol.\\(\%)} &\makecell{Avg.\\eq.\\viol.\\(\%)}  &\makecell{Avg.\\comp.\\time\\(sec)}\\
        \midrule\midrule
	    ED  &\makecell[l]{\multirow{2}{*}{\makecell[l]{Convex\\Opt.}}}& \makecell{183.01} & 0.00 & 0.00 & $0.00_{\pm 0.00}$&\makecell{93.01} & 0.00 & 0.00 & $0.00_{\pm 0.00}$&
        479.69 & 0.00 & 0.00 & $0.00_{\pm0.00}$ \\
        
		DC-OPF  &  & \makecell{183.01} & 0.00 & 0.00 & $0.00_{\pm 0.00}$ & \makecell{93.16} & 0.00 & 0.00 & $\text{0.00}_{\pm0.00}$
        & 522.94 & 0.00 & 0.00 & $0.01_{\pm0.00}$\\
        \midrule
        \multirow{6}{*}{\makecell[l]{DC-OTS}} 
        & \makecell[l]{MIP-M} & 183.01  & 0.00 & 0.00 & \textcolor{red}{$0.36_{\pm0.45}$} & 93.01  & 0.00 & 0.00 & \textcolor{red}{$2.27_{\pm1.72}$}& - & - & - & \textcolor{red}{NA.}\\
        \cmidrule{2-14}
        & \makecell[l]{SL} & \textbf{183.01} & 0.00 & 0.00 & $\text{0.00}_{\pm0.00}$& 93.03 & 0.00 & 0.00 & $\text{0.00}_{\pm0.00}$ &-&-&-&\textcolor{red}{NA.}\\
         & \makecell[l]{LDF} & \makecell{163.46} & \makecell{\textcolor{red}{22.50}} &\makecell{\textcolor{red}{100.00}} &$\text{0.00}_{\pm0.00}$& \makecell{4.22} & \makecell{\textcolor{red}{14.52}} &\makecell{\textcolor{red}{100.00}} &$\text{0.00}_{\pm0.00}$
        &194.48 & \textcolor{red}{32.36} & \textcolor{red}{100.00} & $0.00_{\pm0.00}$\\
        & \makecell[l]{LDF+PP} & - & \makecell{\textcolor{black}{0.00}} & \makecell{\textcolor{red}{20.54}} & $\text{0.00}_{\pm0.00}$& - & \makecell{\textcolor{black}{0.00}} & \makecell{\textcolor{red}{38.82}} & $\text{0.00}_{\pm0.00}$ & - & \textcolor{black}{0.00} & \textcolor{red}{38.65} & $0.00_{\pm0.00}$\\
		&\makecell[l]{\textbf{DA-DNN}} & \makecell{\textbf{183.01}} & 0.00 & 0.00 & $\text{0.00}_{\pm0.00}$& \makecell{\textbf{93.02}} & 0.00 & 0.00 & $\text{0.00}_{\pm0.00}$&
        \textbf{514.00} & 0.00 & 0.00 & $0.01_{\pm0.00}$\\
		\bottomrule
	\end{tabular}
    \captionsetup{justification=raggedright, singlelinecheck=false, textfont=normalfont}
    \caption*{ -: cases where the corresponding quantity cannot be obtained;  $\mu_{\pm\sigma}$: mean and standard deviation of the computation time; NA.: Not available. The generation cost of LDF+PP is not determined since the predicted topologies render DC-OPF infeasible for all test instances. Infeasibility of LDF+PP is detected by the presence of non-zero slack variables in the DC-OPF.}\vspace{-7mm}
\end{table*}

\begin{itemize}
    \item \textbf{ED (Economic Dispatch)}: 
    A baseline that ignores all network constraints and optimizes only the generation cost subject to generator limits. The solution of ED provides a theoretical lower bound on the operating cost.

    \item \textbf{DC-OPF}:
    A convex optimization problem that represents the cost of operating the grid without switching control.

    \item \textbf{DC-OTS (MIP-M)}:
    Big-M formulations of DC-OTS\footnote{We choose the Big‑M formulation of DC‑OTS as a baseline, rather than the original MIQCP formulation, which is known to perform poorly \cite{fattahi2018bound}.}. Let $\Xi_\text{OTS}^{\text{BM}}:=
\{\mathbf{p}_g,\bm{\theta},\mathbf{z},\mathbf{p}_l\}$. Then, we have
\begingroup
\allowdisplaybreaks
\begin{subequations}\label{Eq:DC-OTS-BigM}
\begin{align}
&
\min_{\Xi_\text{OTS}^{\text{BM}}}
C(\mathbf{p}_g)
\label{Eq:BM_Obj}
\\
&
-\mathbf{U}_l \odot \mathbf{z}
\;\le\;
\mathbf{p}_l
\;\le\;
\mathbf{U}_l \odot \mathbf{z}
\label{Eq:BM_OnOff}
\\
&
-\mathbf{M}_l \odot (\mathbf{1}-\mathbf{z})
\;\le\;
\mathbf{p}_l
-
\text{diag}(\mathbf{b}_l)\mathbf{C}\bm{\theta}
\;\le\;
\mathbf{M}_l \odot (\mathbf{1}-\mathbf{z})
\label{Eq:BM_Linking}
\\
&
\text{(\ref{Eq:OTS_PF})},\text{(\ref{Eq:OTS_gen_limits})--(\ref{Eq:OTS_slack})}.
\label{Eq:BM_Other}
\end{align}
\end{subequations}
\endgroup
where $\mathbf{z}\in\{0,1\}^{N_l}$. Here, $\mathbf{M}_l \in \mathbb{R}^{N_l}$ is defined as $\mathbf{M}_l 
:= 
2|\mathbf{b}_l|\overline{\bm{\theta}}$ and $\mathbf{U}_l := \min\!\left(\overline{\mathbf{p}}_{l}, \mathbf{M}_{l} \right)$, where $|\cdot|$ and $\min$ are applied element-wise.

    \item \textbf{SL (Supervised Learning)}:
   Supervised model first predicts the binary line-status vector $\mathbf{z} \in \{0,1\}^{N_l}$ from the load vector $\mathbf{p}_d$, as done in \cite{bugaje2023real}. With the predicted topology, DC-OPF problem is solved to obtain the corresponding  dispatch $(\mathbf{p}_g, \boldsymbol{\theta})$. This method is applicable only when sufficient OTS labels are available for training.

    \item \textbf{LDF (Lagrange-Dual Formulation)}: Unsupervised model that maps $\mathbf{p}_d$ directly to $\mathbf{p}_g$, $\boldsymbol{\theta}$, and $\mathbf{z}$. The model is trained to minimize the Lagrange function without explicitly enforcing physical constraints \cite{fioretto2020lagrangian}. The predicted operating point may violate multiple physical constraints and result in infeasible solutions.

    \item \textbf{LDF+PP (LDF + Post-Processing):}
    LDF is used to obtain the line status $\mathbf{z}$. After that, a single DC-OPF is solved to obtain the generator dispatch and voltage angles. This baseline evaluates whether feasible operating points can be obtained by solving DC-OPF under the predicted line status, which is similar to \cite{pan2020deepopf}, even though LDF does not explicitly enforce physical constraints.
\end{itemize}

\subsubsection{Implementation Details}

All learning-based methods are implemented in PyTorch. We use three layers of fully connected neural network for the line switching network. The hidden dimension is set to 128 for the IEEE 73 bus system and IEEE 118 bus system, and 256 for the IEEE 300 bus system. We employ the exponential linear unit (ELU) as the activation function for all hidden layers \cite{clevert2015fast}. The dropout rate of each layer is 0.2 for 73 and 118 bus systems, and 0.5 for 300 bus system \cite{srivastava2014dropout}. We use a sigmoid function for the final line status output layer of DA-DNN to ensure $\mathbf{\hat{z}} \in [0,1]^{N_\ell}$.

The learning-based baselines (SL, LDF, and LDF+PP) and the proposed DA-DNN are trained using the AdamW optimizer~\cite{loshchilov2017decoupled} with a learning rate of $10^{-4}$ for the IEEE 73 and 118 bus systems and $10^{-3}$ for the IEEE 300-bus system, with a weight decay of $10^{-2}$. 
We use a mini-batch size of 50 and train for up to 40 epochs for the IEEE 73- and 118-bus systems and 80 epochs for the IEEE 300-bus system. The DC-OPF layer in DA-DNN is implemented via CVXPYLayer~\cite{agrawal2019differentiable} and solved using the Clarabel solver~\cite{Clarabel_2024} during training. For the inference phase, all optimization problems are solved using the Gurobi Optimizer v13.0.0~\cite{gurobi}, except for the BLP formulation, which is solved using Ipopt (MA57)~\cite{wachter2006implementation} v3.14.19, Knitro v15.1.0 with and without 100 multi-starts~\cite{byrd2006knitro}. For the mixed-integer DC-OTS formulation, we impose a time limit of 15 minutes, use 64 threads, and set MIP optimality gap of $10^{-5}$. We set the binary regularization weight $\alpha = 10$, which results in the lowest generation cost among ${0, 1, 10, \ldots, 10^{5}}$.

\subsection{Overall Performance Comparisons}

The overall performance comparisons are summarized in Table~\ref{table:118_73_bus}.
We discuss the results separately for the IEEE 73-bus system for non-binding network condition and IEEE 118-bus and 300-bus systems for binding network condition.

\subsubsection{IEEE 73-bus system}

IEEE 73-bus system operates in a non-binding condition, where network constraints are rarely active under the tested loading conditions. Thus, ED, DC-OPF, and DC-OTS yield identical generation costs across all scenarios (\$183.01k), indicating that transmission switching does not provide additional economic benefits in this system.

In this setting, the proposed DA-DNN successfully reproduces the DC-OPF and DC-OTS solutions without inducing unnecessary topology changes. This behavior serves as an important sanity check: when transmission switching is not beneficial, DA-DNN does not introduce spurious line openings and converges to the same operating point as conventional optimization-based methods.

In contrast, learning-based baselines exhibit clear failure modes even in this simple operating condition. The LDF method reports the lowest generation cost (\$163.46k); however, these values arise from operating points that violate all equality constraints and, in some cases, inequality constraints (22.50\%), since physical constraints are only encouraged via penalty terms. Although post-processing (LDF+PP) reduces equality violations (20.54\%) by solving a DC-OPF after predicting line statuses, infeasibility can still occur because the predicted topology may render the subsequent DC-OPF infeasible. SL attains the same generation cost as DA-DNN in this non-binding case, but requires pre-solved DC-OTS labels, which becomes prohibitive at scale (e.g., the 300-bus system). These suggest that explicitly incorporating physical constraints within training improves feasibility.

\subsubsection{IEEE 118-bus system}

The IEEE 118-bus system represents a binding operating condition in which network constraints become active and transmission switching yields measurable economic benefits. As shown in Table~\ref{table:118_73_bus}, DA-DNN attains \$93.02k, which is lower than DC-OPF (\$93.16k) and within 0.01\% of the DC-OTS optimum (MIP-M) while maintaining zero constraint violations. 

Beyond generation cost, another key distinction is computation time. In the IEEE 118-bus system, DA-DNN shows a cost within 0.01\% of MIP-M, while MIP-M exhibits a higher and less predictable runtime ($2.27_{\pm1.72}$\,sec). In contrast, DA-DNN requires only one forward pass and a single DC-OPF, yielding consistently short and predictable time ($0.00_{\pm0.00}$\,sec).

Learning-based baselines exhibit additional limitations. In Table~{\ref{table:118_73_bus}}, SL achieves competitive cost but depends on pre-solved DC-OTS labels, which require solving numerous MIP problems offline and limits scalability. Meanwhile, LDF and LDF+PP continue to violate physical constraints and fail to obtain feasible operating points, consistent with their behavior in the 73-bus case. Thus, DA-DNN attains near-optimal cost while requiring substantially less and more predictable computation time than MIP solvers.

\subsubsection{IEEE 300-bus system}

Unlike the IEEE 73-bus and 118-bus systems, which allow either exact or near-exact solutions of DC-OTS within practical time limits, the IEEE 300-bus system represents a substantially larger case where mixed-integer DC-OTS becomes computationally intractable. The results for this system are summarized in the right columns of Table~\ref{table:118_73_bus}. To enable evaluation over the full test set, the MIP-M baseline is solved with a 15-minute time limit per instance \footnote{We also attempted to solve the DC-OTS with a 48-hour time limit per instance; however, the solver did not converge to an optimal solution within this limit for the IEEE 300-bus system.}.

\begin{figure}[t]
	\centering
\includegraphics[width=1\columnwidth]{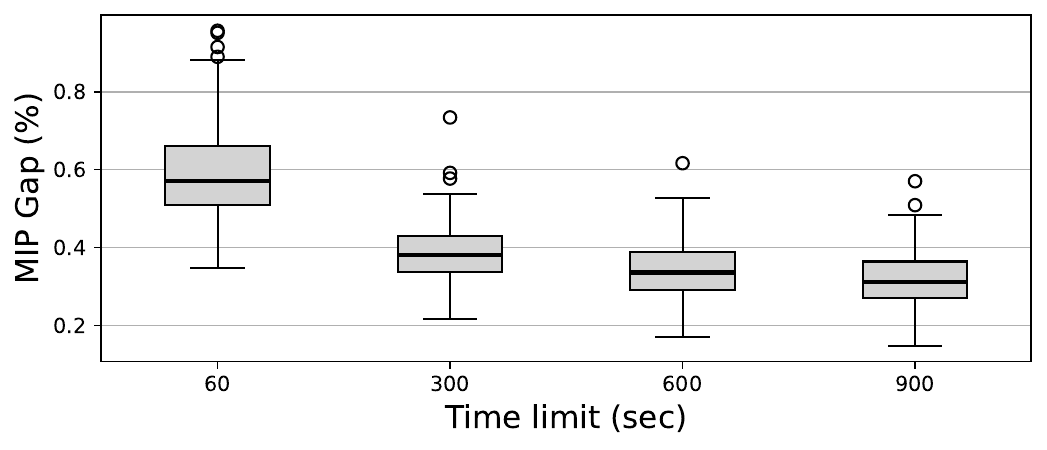}\vspace{-2mm}
	\caption{\small Time varying MIP gap in solving 300 bus system (MIP-M).}
	\label{fig:box_plot}\vspace{-5mm}
\end{figure}

As shown in Table~\ref{table:118_73_bus}, OTS provides clear economic benefits in this system. The proposed DA-DNN achieves an average generation cost of \$514.00k with zero equality and inequality violations, which is 1.70\% lower than  DC-OPF (\$522.94k). In contrast, the time-limited MIP-M baseline struggles to substantially reduce the MIP gap across instances. Fig.~\ref{fig:box_plot} presents the distribution of MIP gaps under the imposed time limit, where noticeable variability and non-negligible gaps remain. These results reflect the difficulty of tightening the bounds in large-scale DC-OTS within practical time limits. Despite this challenge, DA-DNN produces feasible solutions with millisecond-level computation time ($0.01_{\pm0.00}$sec) while maintaining competitive economic performance. 

\begin{table}[t]
\centering
\captionsetup{justification=centering, labelsep=period, font=footnotesize, textfont=sc}
\caption{Comparison in solving DC-OTS using continuous relaxation (Bilinear programming; BLP) and binarization. We use 1000 samples for 300 bus system.}
\label{table:results_cont_relax}

\setlength{\tabcolsep}{4pt}
\renewcommand{\arraystretch}{1.15}

\begin{tabular}{c|c|cc|cc}
\toprule
\multirow{4}{*}{\makecell{Test\\system}} & \multirow{4}{*}{\makecell{Solver\\(method)}} &
\multicolumn{2}{c|}{\makecell{Binary regularization\\deactivated ($\alpha = 0$)}} & \multicolumn{2}{c}{\makecell{Binary regularization\\activated ($\alpha = 10$)}} \\
\cmidrule(lr){3-4} \cmidrule(lr){5-6}
& &\multicolumn{1}{c}{\makecell{Feasible\\ratio (\%)}}
&\multicolumn{1}{c|}{\makecell{Avg. comp.\\time (sec)}}
&\multicolumn{1}{c}{\makecell{Feasible\\ratio (\%)}}
&\multicolumn{1}{c}{\makecell{Avg. comp.\\time (sec)}} \\
\midrule\midrule
\multirow{6.5}{*}{\makecell{IEEE\\73 bus\\system}}&\makecell[l]{Ipopt}   & 100.0 & $0.01_{\pm0.03}$  & 100.0 & $0.01_{\pm0.03}$ \\
&\makecell[l]{Ipopt+W.S.}   & 100.0 & $0.01_{\pm0.02}$ & 100.0 &$0.01_{\pm0.03}$\\
&\makecell[l]{Knitro} & 98.2 & $0.02_{\pm0.04}$ & 100.0  & $0.03_{\pm0.05}$ \\
&\makecell[l]{Knitro+W.S.}  & 100.0 & $0.02_{\pm0.04}$ & 100.0 &$0.01_{\pm0.03}$\\
&\makecell[l]{Knitro+M.S.} & 80.0 & $1.18_{\pm3.45}$ & 98.2 & $1.52_{\pm2.51}$ \\
\cmidrule{2-6}
&\makecell[l]{\textbf{DA-DNN}}  & \textbf{100.0}  & $\mathbf{0.00_{\pm0.00}}$ & \textbf{100.0}  & $\mathbf{0.00_{\pm0.00}}$ \\
\midrule
\multirow{6.5}{*}{\makecell{IEEE\\118 bus\\system}}&\makecell[l]{Ipopt}   & 98.6 & $0.03_{\pm0.03}$  & 100.0 & $0.07_{\pm0.04}$ \\
&\makecell[l]{Ipopt+W.S.}   & 98.7 & $0.02_{\pm0.02}$ & 100.0 &$0.07_{\pm0.04}$\\
&\makecell[l]{Knitro}  & 84.9 & $0.04_{\pm0.04}$ & 84.8  & $0.05_{\pm0.04}$ \\
&\makecell[l]{Knitro+W.S.}  & 94.9  & $0.03_{\pm0.05}$ & 100.0 &$0.03_{\pm0.04}$\\
&\makecell[l]{Knitro+M.S.}  & 59.5  & $3.31_{\pm10.27}$ & 81.7  & $3.82_{\pm11.75}$ \\
\cmidrule{2-6}
&\makecell[l]{\textbf{DA-DNN}}  & \textbf{100.0} & $\mathbf{0.00_{\pm0.00}}$ & \textbf{100.0}   & $\mathbf{0.00_{\pm0.00}}$ \\
\midrule
\multirow{6.5}{*}{\makecell{IEEE\\300 bus\\system}}&\makecell[l]{Ipopt}   & 0.0  & $4.16_{\pm4.47}$ & 0.0  &$1.31_{\pm0.70}$\\
&\makecell[l]{Ipopt+W.S.}   & 0.2 & $5.04_{\pm5.20}$ & 0.0  &$1.66_{\pm0.88}$\\
&\makecell[l]{Knitro}  & 0.0  & $1.73_{\pm7.76}$ & 0.0   & $2.33_{\pm10.20}$\\
&\makecell[l]{Knitro+W.S.}  & 0.0  & $1.99_{\pm6.73}$ & 0.2  & $1.23_{\pm0.77}$\\
&\makecell[l]{Knitro+M.S.}  & 0.0  & $35.20_{\pm42.09}$ & 0.0  & $40.43_{\pm46.65}$\\
\cmidrule{2-6}
&\makecell[l]{\textbf{DA-DNN}}  & \textbf{100.0} & $\mathbf{0.01_{\pm0.00}}$ & \textbf{100.0}  & $\mathbf{0.01_{\pm0.00}}$ \\
\bottomrule
\end{tabular}
\captionsetup{justification=raggedright, singlelinecheck=false, textfont=normalfont}
    \caption*{Feasible ratio: Feasible ratio of test instances after binarization; Avg. comp. time: Computation time in solving BLP; W.S.: Warm-start with DC-OPF solution ($\mathbf{p}_g$ and $\bm{\theta}$) and $\mathbf{z} = \sigma(9)$; M.S.: 100 multi-starts.
    }\vspace{-6mm}
\end{table}

\subsection{Overall comparisons with Solvers in Solving BLP}
To further evaluate the effectiveness of DA-DNN against solver-based baselines, we compare it with a bilinear programming (BLP) relaxation of DC-OTS, i.e., (\ref{Eq:DC-OTS}) with continuous $\mathbf{z}\in[0,1]^{N_l}$. This comparison is relevant because BLP directly solves the optimization problem used to generate $\frac{\partial C(\hat{\mathbf{p}}_g^*)}{\partial \hat{\mathbf{z}}}$ in DA-DNN. Specifically, we solve the BLP using Ipopt, Ipopt+W.S. (warm-start from the DC-OPF solution with $\mathbf{z}=0.9999$), Knitro, Knitro+W.S., and Knitro+M.S. (100 random initializations), followed by the same binarization in (\ref{eq:binarization}) and a DC-OPF solve. The results are summarized in Table~\ref{table:results_cont_relax}. Computations were performed on CPU servers equipped with dual Intel Xeon Platinum 8452Y+ 2.0~GHz processors (72 cores) and 250~GB memory.

For the IEEE 73-bus system, solver-based methods attain high feasible ratios after binarization. Ipopt with and without warm-start achieves 100\% feasibility, and Knitro achieves 98.2\% without warm-start and 100\% with warm-start. This is consistent with the non-binding operating condition of the IEEE 73-bus system, where network constraints are rarely active, as shown by the nearly identical generation costs to DC-OPF in Table~\ref{table:118_73_bus}. In contrast, multi-start, which initializes from random points, reduces feasibility to 80.0\% under $\alpha=0$. Activating binary regularization improves the multi-start feasible ratio to 98.2\%. The feasible ratio becomes increasingly dependent on warm-start and the regularization in the IEEE 118-bus system. For example, without warm-start, Knitro achieves 84.9\% feasibility, and multi-start further reduces it to 59.5\%. By contrast, Knitro with warm-start raises feasibility to 94.9\%, demonstrating the importance of starting from a physically meaningful operating point. Moreover, binary regularization increases the multi-start feasible ratio from 59.5\% to 81.7\%. Overall, the feasibility in the IEEE 118-bus system is more sensitive to warm-start and binary regularization than in the IEEE 73-bus case, since it operates closer to its network limits.

This limitation becomes even more severe in the IEEE 300-bus system, where all solver-based methods yield 0.0$-$0.2\% feasible ratio after binarization, regardless of initialization strategy or regularization, even though the continuous BLP remains feasible and converges within a few seconds per instance. These results indicate that binarization is particularly problematic in the IEEE 300-bus system: thresholding can switch off critical transmission lines, disconnect the network, and render the post-binarization DC-OPF infeasible.

By contrast, DA-DNN achieves a 100\% feasible ratio across all three systems, including the 300-bus case, with inference time comparable to a single DC-OPF solve ($0.01_{\pm0.00}$\,sec). Since DA-DNN is initialized at the DC-OPF topology and trained by minimizing the expected objective over many operating scenarios, it primarily reduces line-status values only for lines that are consistently beneficial to switch off across the training set. As a result, more lines remain in service, keeping more transmission lines available, and improving post-binarization feasibility. By comparison, the solver-based BLP baselines optimize each instance independently and can rely on fragile fractional couplings that collapse after hard thresholding in the tightly constrained 300-bus system. Results which support this conclusion are provided in the following subsection.

\subsection{Impact of Binary Regularization}

\begin{figure}[t]
     \centering
     \begin{subfigure}[b]{1\columnwidth}
     \centering
         \includegraphics[width=\columnwidth]{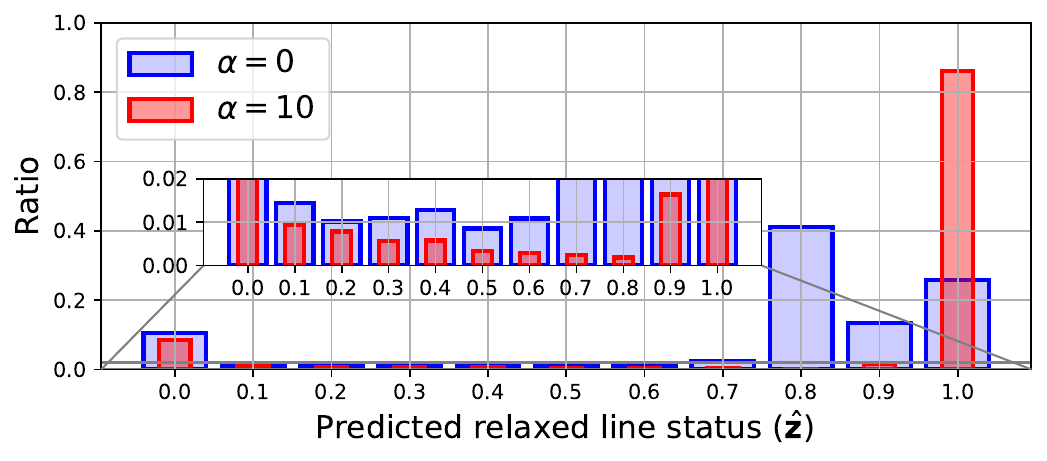}\vspace{-2mm}
         \caption{\small Ipopt+W.S. (MA57).}
         \label{fig:varying_alpha_ipopt}
     \end{subfigure}\vspace{-0.5mm}
     \begin{subfigure}[b]{1\columnwidth}
     \centering
         \includegraphics[width=\columnwidth]{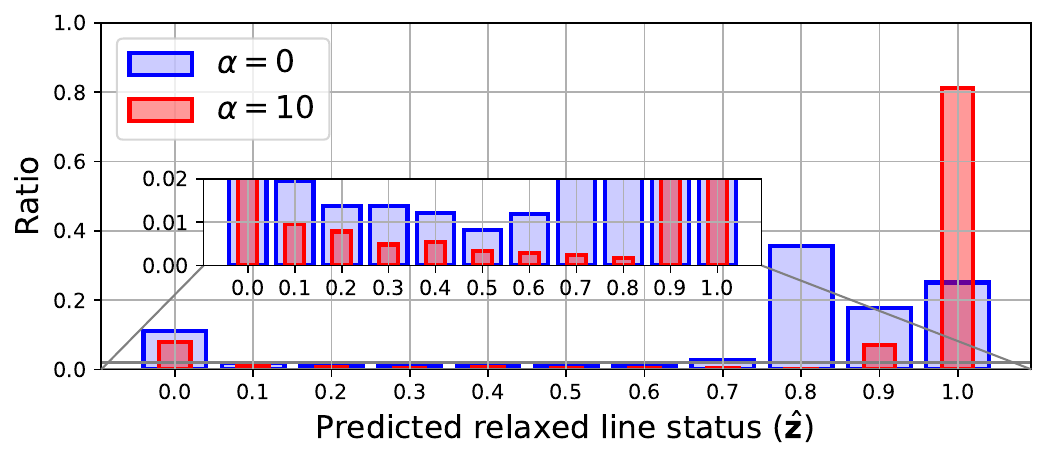}\vspace{-2mm}
         \caption{\small Knitro+W.S.}
         \label{fig:varying_alpha_Knitro}
     \end{subfigure}\vspace{-0.5mm}
     \begin{subfigure}[b]{1\columnwidth}
     \centering
         \includegraphics[width=\columnwidth]{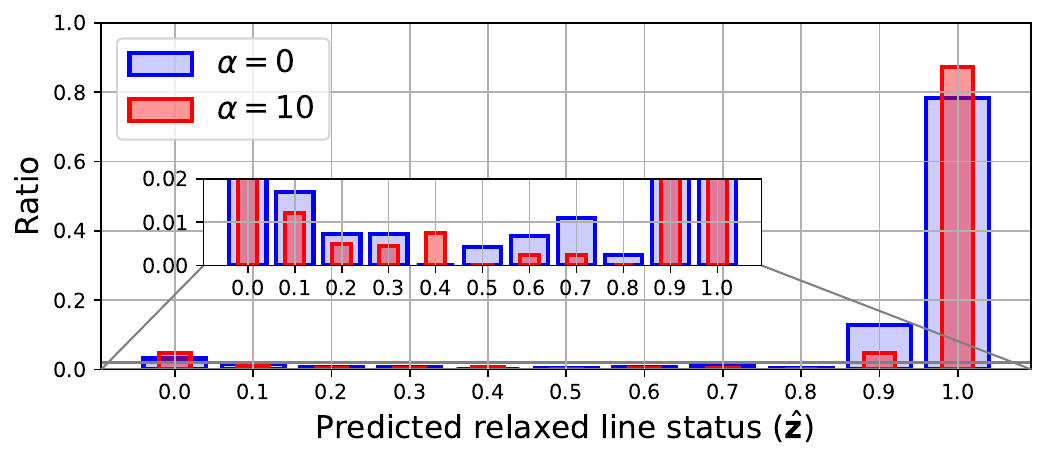}\vspace{-2mm}
         \caption{\small DA-DNN.}
         \label{fig:varying_alpha_2}
     \end{subfigure}\vspace{-1mm}
     \caption{\small Distribution of predicted relaxed line-status $\hat{\mathbf{z}}$ for $\alpha = 0$ (\textcolor{blue}{blue}, wider bars) and $\alpha = 10$ (\textcolor{red}{red}, narrower bars) in 300 bus system. Bars show the empirical ratio per bin: 0.0,...,0.9 denote intervals $[0.0,0.1)$,...,$[0.8, 0.9)$,$[0.9,\sigma(9))$ and $1.0$ is $\hat{z}_l\in[\sigma(9), 1)$.}
	\label{fig:varying_alpha}\vspace{-4mm}
\end{figure}

Fig.~\ref{fig:varying_alpha} compares the distributions of $\hat{\mathbf{z}}$ for $\alpha=10$ and $\alpha=0$ in Ipopt-W.S., Knitro-W.S., and DA-DNN. Binary regularization pushes the relaxed line-status values toward the extremes (0 or 1) for both the solvers and DA-DNN. In particular, activating regularization increases the predicted line status value near $\hat{z}_i=1$ and decreases the density in intermediate bins, indicating that ambiguous decisions are reduced. However, the magnitude and pattern of this shift differ across methods. For the solver-based solutions in Fig.~\ref{fig:varying_alpha_ipopt} and Fig.~\ref{fig:varying_alpha_Knitro}, the distribution shows a larger reduction in the 0.8$-$0.9 bins and a strong increase in $\hat{z}_i\approx 1$. 

By contrast, DA-DNN achieves a more concentrated distribution near $\hat{z}_i=1$ with a smaller fraction near $\hat{z}_i\approx 0$, indicating that the model switches off fewer lines and keeps more lines in service, as shown in Fig.~\ref{fig:varying_alpha_2}. This is because, unlike the solver-based approach that optimizes the BLP independently for each single instance, DA-DNN is trained to minimize the expected loss over a training set. As a result, DA-DNN tends to reduce line-status values only when switching off a line provides a consistent benefit across the training set, while keeping the remaining lines decisively on. In addition, our DC-OPF-based initialization starts the training process from an all-lines-in-service operating point, which encourages the model to maintain lines in-service unless there are clear benefits that switching is beneficial. Consequently, the binarized topologies keep more line flows available and are more likely to remain feasible after thresholding compared with solver-based baselines.

\subsection{Impact of the Weight and Bias Initialization}

\begin{figure}[t]
	\centering
\includegraphics[width=1\columnwidth]{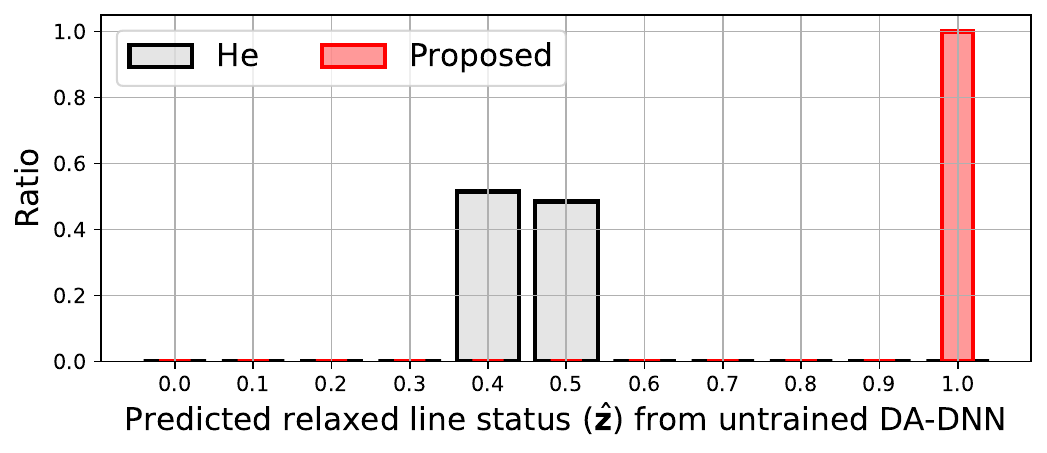}\vspace{-2mm}
	\caption{\small Distribution of the predicted relaxed line status values from untrained DA-DNN for 300 bus system with different weight and bias initialization (He: black, wider bars; Proposed: \textcolor{red}{red}, narrower bars). Bars show the empirical ratio per bin: 0.0,...,0.9 denote intervals $[0.0,0.1)$,...,$[0.8, 0.9)$,$[0.9,\sigma(9))$ and $1.0$ is $\hat{z}_l\in[\sigma(9), 1)$.}
	\label{fig:weight_histogram_2}\vspace{-5mm}
\end{figure}

\begin{figure*}[t]
     \centering
     \begin{subfigure}[b]{0.68\columnwidth}
     \centering
         \includegraphics[width=\columnwidth]{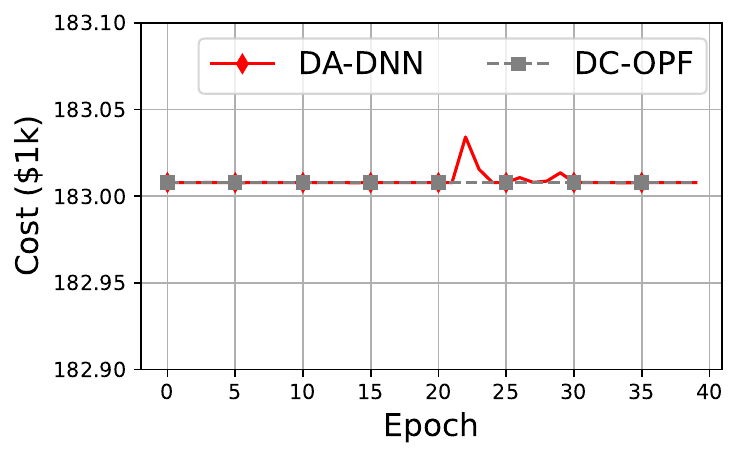}\vspace{-2mm}
         \caption{\small IEEE 73 bus system.}
         \label{fig:case73}
     \end{subfigure}
     \begin{subfigure}[b]{0.67\columnwidth}
     \centering
         \includegraphics[width=\columnwidth]{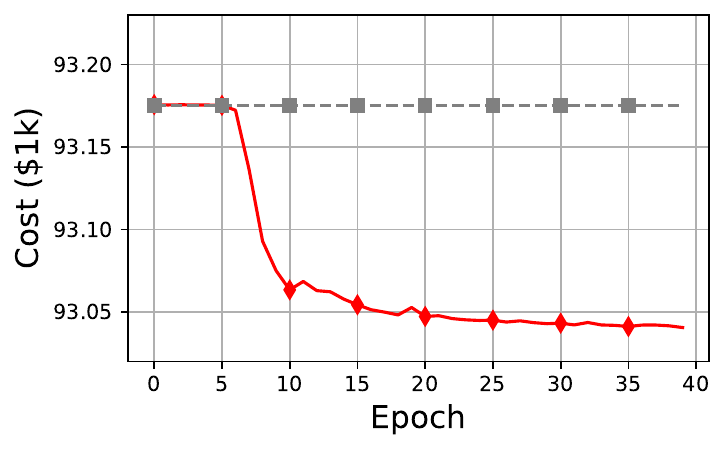}\vspace{-2mm}
         \caption{\small IEEE 118 bus system.}
         \label{fig:case118}
     \end{subfigure}
     \begin{subfigure}[b]{0.65\columnwidth}
     \centering
         \includegraphics[width=\columnwidth]{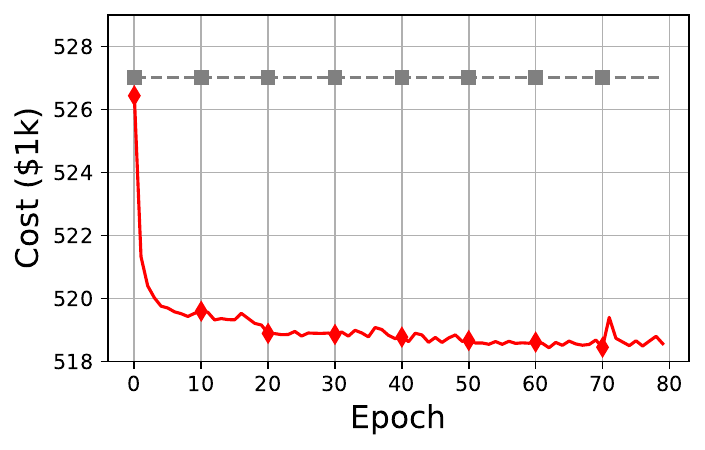}\vspace{-2mm}
         \caption{\small IEEE 300 bus system.}
         \label{fig:case300}
     \end{subfigure}\vspace{-2mm}
     \caption{\small Training curve in each test system.}
     \label{fig:loss_function}\vspace{-6mm}
\end{figure*}

As discussed above, DA-DNN is initialized to start training in a stable and typically feasible region near the DC-OPF topology. To demonstrate the effectiveness of the proposed method, we compare the proposed initialization with standard He initialization~\cite{he2015delving}. In our scheme, the last-layer parameters are set to $\mathbf W_{\mathrm{init}}^\text{last}=\mathbf 0$ and $\mathbf b_{\mathrm{init}}^\text{last}=9$, yielding $\hat{\mathbf z}\approx\mathbf 1$ at the first forward pass. This corresponds to the all-lines-in-service topology and makes the embedded optimization approximately equivalent to DC-OPF at epoch~0. As shown in Fig.~\ref{fig:weight_histogram_2}, the proposed initialization concentrates the relaxed line-status values near one, whereas He initialization yields ambiguous switching values near $0.5$, increasing the likelihood of infeasible initial topologies. Also, as can be seen in Fig.~\ref{fig:loss_function}, the proposed initialization starts from a generation cost identical to DC-OPF at epoch~0 across all test systems and shows smooth convergence. In contrast, He initialization hinders the training process, as the initial topologies lead to infeasibility.

\subsection{Inference with Untrained Line Flow Limits}

In this subsection, we verify the generalization ability and adaptability of the proposed method. For this, DA-DNN is trained under nominal static line ratings (SLR) corresponding to 100\% of line flow limits. However, transmission capacities often deviate from their static values due to ambient and seasonal variations, and the deployment of dynamic line ratings (DLRs). In most real-world DLR applications, the line flow limits are higher than the SLR, since cases where the DLR falls below the SLR occur much less frequently \cite{hou2020research}.

\begin{figure}[t]
	\centering
\includegraphics[width=1\columnwidth]{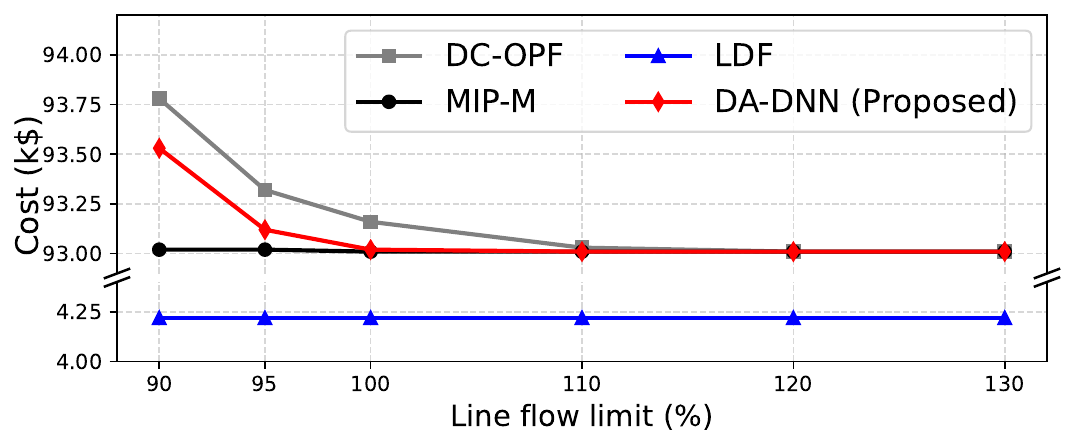}\vspace{-2mm}
	\caption{\small Generation cost comparisons with varying line flow limits. LDF+PP Cost is not shown since it is infeasible for all test instances.}
	\label{fig:varying_lineflows}\vspace{-6mm}
\end{figure}

To examine this practical scenario, we evaluate the generation cost of solver-based (MIP-M) and unsupervised learning-based (LDF and DA-DNN) methods under a wide range of untrained line flow limits between 90\% and 130\%. As shown in Fig.~\ref{fig:varying_lineflows}, DA-DNN closely tracks the behavior of DC-OPF and DC-OTS as the line flow limits vary. For higher limits, which correspond to conditions commonly observed in DLR-enabled systems, DA-DNN achieves costs nearly identical to the optimal OTS solution. Although the predicted topology remains unchanged under untrained line-flow limits, the embedded DC-OPF layer recomputes the dispatch using the updated constraints at inference time. Therefore, when the limits change, the feasible region of the DC-OPF changes accordingly, and the resulting dispatch reflects the modified constraints without retraining. In contrast, the LDF method produces nearly identical outputs across all limit settings because it maps the load vector directly to dispatch without incorporating network constraints. When line flow limits are more restrictive than those used during training, DA-DNN exhibits a moderate increase in generation cost due to the reduced feasible region, while remaining feasible and consistently outperforming DC-OPF.

\subsection{Feasibility recovery from contingency}
\label{subsec:contingency_recovery}

Finally, we evaluate DA-DNN as a post-contingency feasibility recovery method, where corrective switching and redispatch are allowed after a contingency \cite{sheikh2019multiobjective}. 
Let $\Xi_\text{COTS}=\{\Delta\mathbf{p}_g,\bm{\theta},\mathbf{z}\}$ denote the decision variables for corrective DC-OTS, where $\Delta\mathbf{p}_g$ is the redispatch from the dispatch $\mathbf{p}_g^*$, $\bm{\theta}$ is the voltage angle, and $\mathbf{z}$ is the binary line-status vector. Then, the corrective DC-OTS is formulated as follows:
\begingroup
\allowdisplaybreaks
\begin{subequations}\label{Eq:DC-OTS}
\begin{align}
&
\min_{\Xi_\text{COTS}}
C(\mathbf{p}_g^* + \Delta\mathbf{p}_g) + \beta \sum_{l\in\mathcal{L}_{sw}}(1-z_l)
\label{Eq:corr_OTS_Obj}\\
&
\mathbf{M}(\mathbf{p}_g^* + \Delta\mathbf{p}_g) - \mathbf{p}_d = \mathbf{C}^\intercal \big(\text{diag}(\mathbf{z}\odot\mathbf{b}_l)\mathbf{C}\bm{\theta}\big),\label{Eq:corr_OTS_PF}\\
&
\underline{\mathbf{p}}_g - \mathbf{p}_g^* \leq \Delta\mathbf{p}_g \leq \overline{\mathbf{p}}_g - \mathbf{p}_g^*,\label{Eq:corr_OTS_gen_limits}\\
&\text{(\ref{Eq:OTS_line_limits}), (\ref{Eq:OTS_angle_limits}), (\ref{Eq:OTS_slack})}.
\end{align}
\end{subequations}
\endgroup
\noindent
Here, we set $\beta=50$ to discourage unnecessary switching, as done in \cite{flores2020alternative}. $\mathcal{L}_{sw}$ denotes the set of switchable lines.

For each test system, we first identify the line outage scenario that causes the most severe overload under the PGLib load profiles, e.g., IEEE 73-bus (lines 66 and 67 out), IEEE 118-bus (lines 104 and 105 out), and IEEE 300-bus (line 214 out). DA-DNN is trained to solve DC-OTS on the modified test system, where the corresponding lines are permanently removed. Thus, the embedded DC-OPF layer is constructed on the contingency topology. Note that due to the increased complexity of the post-outage test system, we use a wider switching network (hidden dimension 256) with a dropout rate of 0.5 to improve model capacity.

The feasibility recovery results are summarized in Table~\ref{table:contingency}. We compare DA-DNN with the solver-based baseline (MIP-M) in terms of the amount of corrective switching (measured by the disconnected line ratio) and computation time. In the IEEE 73- and 118-bus systems, DA-DNN tends to open more lines than MIP-M (13.33\% vs. 8.33\% for IEEE 73-bus and 3.76\% vs. 3.23\% for IEEE 118-bus). Despite this increased switching, DA-DNN achieves feasibility recovery extremely fast: it completes in $0.00$~sec for the IEEE 73- and 118-bus systems and $0.01$~sec for the IEEE 300-bus system, making it well suited for fast post-contingency recovery where rapid corrective actions are required. By contrast, MIP-M requires seconds even in these systems ($1.65$~sec for IEEE 73-bus and $7.14$~sec for IEEE 118-bus) and fails to return a solution within the 15-minute time limit for the IEEE 300-bus system. Thus, DA-DNN offers a practical corrective scheme with a computation time comparable to a single DC-OPF, while the solver-based approach becomes difficult to use within tight operational time limits as the system size increases.

\begin{table}[t]
\centering
\captionsetup{justification=centering, labelsep=period, font=footnotesize, textfont=sc}
\caption{Feasibility recovery from contingency.}
\label{table:contingency}

\setlength{\tabcolsep}{5pt}
\renewcommand{\arraystretch}{1.15}

\begin{tabular}{c|c|cc}
\toprule
\makecell{Test system} & \makecell{Method} & \makecell{Disconnected\\line ratio (\%)} & \makecell{Computation\\time (sec)}\\
\midrule\midrule
\multirow{2}{*}{\makecell{IEEE 73 bus system\\(line 66, 67 out)}} &\makecell[l]{MIP-M}   & 8.33 & 1.78\\
&\makecell[l]{DA-DNN}  & 15.00 & 0.00 \\
\midrule
\multirow{2}{*}{\makecell{IEEE 118 bus system\\(line 104, 105 out)}} &\makecell[l]{MIP-M}   & 3.23 & 7.14.\\
&\makecell[l]{DA-DNN}  & 4.84 & 0.00 \\
\midrule
\multirow{2}{*}{\makecell{IEEE 300 bus system\\(line 214 out)}} &\makecell[l]{MIP-M}   & - & NA.\\
&\makecell[l]{DA-DNN}  & 7.30 & 0.01 \\
\bottomrule
\end{tabular}
\captionsetup{justification=raggedright, singlelinecheck=false, textfont=normalfont}
    \caption*{NA.: Not available within 15 minutes.
    }\vspace{-6mm}
\end{table}

\section{Conclusion}

In this paper, we proposed DA-DNN, which combines a line switching network with an embedded differentiable DC-OPF layer to solve DC-OTS in an unsupervised manner while enforcing dispatch feasibility through optimization. To stabilize the training process, we introduced a feasible weight and bias initialization that prevents early infeasibility during training and enables reliable convergence. To improve inference reliability, we incorporated a binary regularization term that reduces ambiguity in relaxed line-status outputs, leading to more stable binarized switchings and improved generation cost performance after thresholding. The case studies show that DA-DNN scales to large networks and provides highly predictable inference time, making it suitable for integration into real-time operational pipelines. Furthermore, DA-DNN generalizes to unseen line-flow limit settings without retraining, demonstrating robustness to time-varying network constraints. Even when DC-OTS is computationally intractable, DA-DNN provides a practical alternative that results in feasible switching decisions with consistent computational performance, enabling real-time post-contingency recovery. For future work, we will extend the framework by replacing the DC-OPF layer with an AC-OPF layer and incorporating additional operational requirements such as uncertainty and stability constraints.

\balance
\bibliographystyle{IEEEtran}
\bibliography{reference.bib}
\end{document}